\documentclass[aps,prx,twocolumn,superscriptaddress]{revtex4-1}  
\usepackage{graphicx}  
\usepackage{dcolumn}   
\usepackage{bm}        
\usepackage{amssymb}   
\usepackage{amsmath}
\usepackage[latin1]{inputenc}
\usepackage{color}

\begin{document}


\title{Two-dimensional superconductivity in a bulk single-crystal}

\author{Q. R. Zhang}
\affiliation{National High Magnetic Field Laboratory, Florida State University, Tallahassee-FL 32310, USA}
\affiliation{Department of Physics and National High Magnetic Field Laboratory, Florida State University, Tallahassee, Florida 32310, USA}
\author{D. Rhodes}
\affiliation{National High Magnetic Field Laboratory, Florida State University, Tallahassee-FL 32310, USA}
\affiliation{Department of Physics and National High Magnetic Field Laboratory, Florida State University, Tallahassee, Florida 32310, USA}
\author{B. Zeng}
\affiliation{National High Magnetic Field Laboratory, Florida State University, Tallahassee-FL 32310, USA}
\author{M. D. Johannes}
\affiliation{Center for Computational Materials Science, Naval Research Laboratory, Washington, DC 20375, USA}
\author{L. Balicas}
\affiliation{National High Magnetic Field Laboratory, Florida State University, Tallahassee-FL 32310, USA}
\date{\today}

\begin{abstract}
Both Nb$_3$Pd$_x$Se$_7$ and Ta$_4$Pd$_3$Te$_{16}$ crystallize in a monoclinic point group while exhibiting superconducting
transition temperatures as high as $T_c\sim 3.5$ and $\sim 4.7 $ K, respectively. Disorder was claimed to lead to the extremely large upper critical fields ($H_{c2}$) observed in related compounds. Despite the presence of disorder and heavier elements, $H_{c2}$s in Ta$_4$Pd$_3$Te$_{16}$ are found to be considerably smaller than those of Nb$_3$Pd$_x$Se$_7$ while displaying an anomalous, non-saturating linear dependence on temperature $T$ for fields along all three crystallographic axes. In contrast, crystals of the latter compound displaying the highest $T_c$s display $H_{c2}\propto (1-T/T_c)^{1/2}$, which in monolayers of transition metal dichalcogenides is claimed to be evidence for an Ising paired superconducting state resulting from strong spin-orbit coupling. This anomalous $T$-dependence indicates that the superconducting state of Nb$_3$Pd$_x$Se$_7$ is quasi-two-dimensional in nature. This is further supported by a nearly divergent anisotropy in upper-critical fields, i.e. $\gamma= H_{c2}^{b}/H_{c2}^{a^{\prime}}$, upon approaching $T_c$. Hence, in Nb$_3$Pd$_x$Se$_7$ the increase of $T_c$ correlates with a marked reduction in electronic dimensionality as observed, for example, in intercalated FeSe. For the Nb compound, Density functional theory (DFT) calculations indicate that an increase in the external field produces an anisotropic orbital response, with especially strong polarization at the Pd sites when the field is perpendicular to their square planar environment.  The field also produces an anisotropic spin moment at both Pd sites.   Therefore, DFT suggests the field-induced pinning of the spin to the lattice as a possible mechanism for decoupling the superconducting planes. Overall, our observations represent further evidence for unconventional superconductivity in the Pd chalcogenides.

\end{abstract}

\maketitle

\section{introduction}

Recently, extremely high superconducting upper-critical fields were observed in ultra thin single-crystals of transition-metal dichalcogenides for magnetic fields
applied along a planar direction \cite{Mak,Iwasa}. In effect, in single-layered NbSe$_2$ and in electric field-induced superconducting MoS$_2$,
upper critical fields exceeding the weak coupling Pauli limiting field, respectively by factors of 6 and 4, were reported \cite{Mak,Iwasa,Iwasa2}. Single atomic layers of both compounds are characterized by lack of inversion symmetry and a strong spin orbit coupling, which couples (or locks) the spins of the carriers along the orbital moments pointing
perpendicularly to the conducting planes. Valley degeneracy and time reversal symmetry imply that the orbital moments and coupled spins, must have opposite polarization between
the valleys at the \emph{K} and the \emph{-K} points. While multi-layers are spin degenerate, in single atomic layers the aforementioned coupling induces a Zeeman-like spin-splitting of the valence and conduction bands as observed at the \emph{K}-point  of their hexagonal Brillouin zone \cite{zhu}. Ising-like superconducting pairing is proposed to occur between carriers (of opposite spin- and valley-polarization) at the \emph{K} and \emph{-K} valleys. This state would be particularly robust against an external in-plane magnetic field due to the spin-valley locking leading to the extremely large values of $H_{c2}$ seen experimentally \cite{Mak,Iwasa}.

Extremely high upper critical fields were also recently reported for a new family of Pd based quasi-one-dimensional transition-metal chalcogenides
containing Nb and Ta \cite{Alan1,Alan2,takagi,kim,jiao,jiao2}.  Although these compounds display relatively low $T_c$s, i.e. ranging from $\sim 7.4$ K to $\sim 1.5$ K
for Nb$_2$Pd$_{0.81}$S$_5$ \cite{Alan1} and Ta$_3$Pd$_3$Te$_{14}$ \cite{jiao2} respectively, they tend to display upper critical fields that greatly exceed the weakly
coupling Pauli limiting field value $H_p \simeq 1.84 T_c$. For example, for  Nb$_2$Pd$_{0.81}$S$_5$ one obtains $H^b_{c2}(T \rightarrow 0 \text{ K}) \sim$ 37 T for fields along its $b-$axis \cite{Alan1}. These compounds are characterized by structural disorder due, for instance, to the off stoichiometry of the Pd atoms. Disorder was proposed to suppress the paramagnetic pair-breaking effect due to strong spin-orbit scattering thus producing their very high $H_{c2}$s \cite{takagi}. A very large spin-orbit coupling parameter was extracted from a fit to the Werthamer-Helfand-Hohenberg formalism of the superconducting to metallic phase-boundary of Nb$_3$Pd$_{x}$Se$_7$ \cite{Alan2}. In Nb$_2$Pd$_x$S$_5$ the isovalent substitution of Pd for the heavier element Pt was found to enhance the ratio of $H_{c2}$ to $T_c$, while this ratio is slightly suppressed for Ni doping \cite{Zhou}. This represents additional evidence indicating that spin-orbit coupling on the Pd sites plays a relevant role for its superconducting properties contributing to its large upper critical fields.

According to Density Functional theory calculations \cite{Alan1,Alan2,Singh1,Singh2,Singh3} the Fermi surface of these compounds tend to be complex being composed of corrugated quasi-one-dimensional sheets and quasi-two-dimensional surfaces as in Nb$_2$Pd$_{x}$S$_5$ \cite{Alan1}, in addition to a complex three dimensional network as in Nb$_3$Pd$_{x}$Se$_7$ \cite{Alan2}. In both compounds the superconducting anisotropy $\gamma$, defined as the ratio of $H^b_{c2}$s for fields along the needle or the \emph{b}-axis with respect to $H_{c2}$s for fields applied along the other crystallographic orientations, is temperature dependent as seen in the Fe pnictide superconductors and is interpreted as evidence for multi-band superconductivity. This conclusion is supported by Ref. \onlinecite{Singh1}, which claims that Ta$_2$PdS$_5$ should be considered as a two-band strong-coupled superconductor in the dirty limit. Ref. \onlinecite{Singh1} argues that the nearly linear dependence of $H_{c2}$ on temperature $T$ displayed by most of these Pd based compounds would correspond to experimental evidence for such scenario. Based on the previous paragraphs one would conclude that disorder, multi-band superconductivity and spin-orbit coupling are the basic  ingredients leading to the very large $H_{c2}$s observed in the Pd based Ta and Nb chalcogenides.

However, scanning tunnelling spectroscopy measurements on Ta$_4$Pd$_3$Te$_{16}$, with a middle point transition at $T_{c}^{\text{mid}} \simeq 4.5$ K \cite{jiao}, reveals evidence for either an anisotropic superconducting $s-$wave gap, i.e. having gap minima, or the possibility of gap nodes (a $d-$wave component) in a multi-band system \cite{HHWen}. In addition, its extreme anisotropy leads to elongated vortices with a core anisotropy of $\xi_{\|b}/ \xi_{\bot b} \approx 2.5$ \cite{HHWen}. Furthermore, low-temperature thermal conductivity measurements in this compound reveal a significant residual electronic term at zero magnetic field which, similarly to the cuprates, increases rapidly as the field increases hence corresponding to evidence for nodes in its gap function \cite{Pan}. This, coupled to the existence of a superconducting dome in its temperature as a function of pressure phase-diagram would, according to Ref. \onlinecite{Pan}, correspond to evidence for an unconventional superconducting state in Ta$_4$Pd$_3$Te$_{16}$.

Here, we report the phase diagram of Nb$_3$Pd$_{x=0.84}$Se$_{7+\delta}$ single-crystals displaying a middle point superconducting transition at $T_c \simeq 3.5$ K which is nearly a factor two higher than the previously reported value. This suggests that this series also displays a superconducting dome as a function of the Pd content. More importantly, for fields along their needle axis these crystals display an anomalous superconducting to metallic phase-boundary with $H^b_{c2}$ displaying a $\propto (1-T/T_c)^{1/2}$ dependence on temperature over the entire $T$ range. This is the functional dependence observed in single-crystals composed of single- or few atomic-layers of transition-metal dichalcogenides for fields applied along a planar direction \cite{Mak,Iwasa,Iwasa2}. Such a dependence leads to a nearly divergent superconducting anisotropy $\gamma$ upon approaching $T_c$ which is consistent with two-dimensional superconductivity. This contrasts with the more conventional phase-diagram extracted from samples displaying lower $T_c$s, indicating a remarkable increase in electronic and hence in superconducting anisotropy and possibly a dependence of the superconducting gap symmetry on Pd stoichiometry. This anomalous $T-$dependence might be attributable to the locking of the spins along a direction perpendicular to the needle axis due to strong spin-orbit (SO) coupling, although our band-structure calculations indicate that SO-coupling is relatively weak for this compound. Furthermore, we observe considerably smaller values of $H_{c2}$ for Ta$_4$Pd$_3$Te$_{16}$, which is also monoclinic and composed of considerably heavier elements, albeit with its $H_{c2}$s displaying an anomalous linear dependence on temperature over the entire $T$ range. For both compounds the absence of saturation in the low temperature values of $H_{c2}$, particularly when the magnetic field exceeds the Pauli limiting value, is at odds with conventional singlet superconductivity. Our observations suggest that Ta$_4$Pd$_3$Te$_{16}$ probably is an orbital limited superconductor over the entire $T$ range. As for Nb$_3$Pd$_{x}$Se$_7$, we argue, based on band structure calculations, that Ising pairing scenario is unlikely to explain our observations. Instead, its non-stoichiometric composition is likely to induce a very weak coupling between superconducting planes leading to rather small superconducting coherence lengths and possibly to the concomitantly high upper critical fields observed by us.
\section{experimental}
\begin{figure}[htp]
\begin{center}
    \includegraphics[width=7.5 cm]{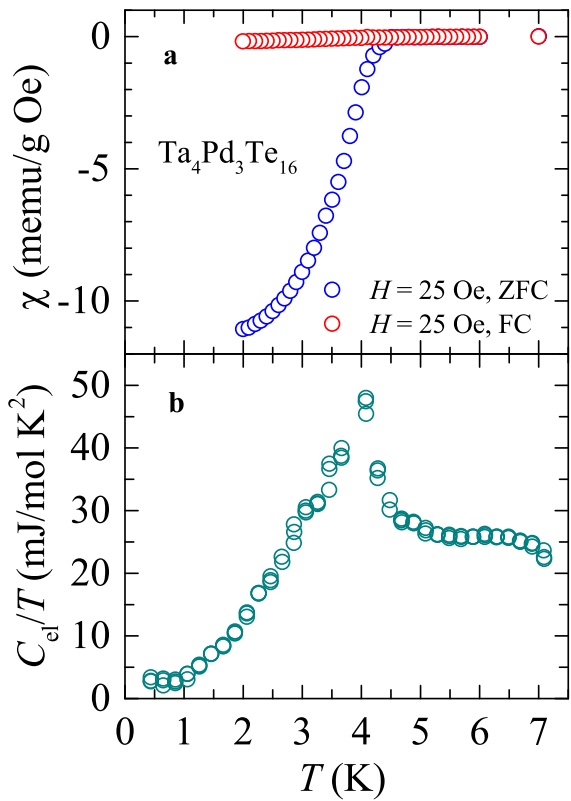}
    \caption{(a)  Magnetic susceptibility $\chi$ as of temperature for several randomly oriented Ta$_{4}$Pd$_{3}$Te$_{16}$ single crystals.  $\chi (T)$ measured under zero-field cooled conditions is depicted by blue makers, while $\chi (T)$ measured under field-cooled conditions is indicated by red markers. The pronounced deviation among both curves observed below $T_c \sim 4.5$ K indicates a bulk superconducting transition. (b) Electronic contribution to the heat capacity $C_{\text{el}}$, normalized by the temperature $T$, as obtained after subtracting a $C \propto T^3$ term (i.e. phonon contribution). A pronounced anomaly is observed at $T_c$.}
    \label{TPTchi}
\end{center}
\end{figure}
\begin{figure*}[htp]
\begin{center}
    \includegraphics[width=12 cm]{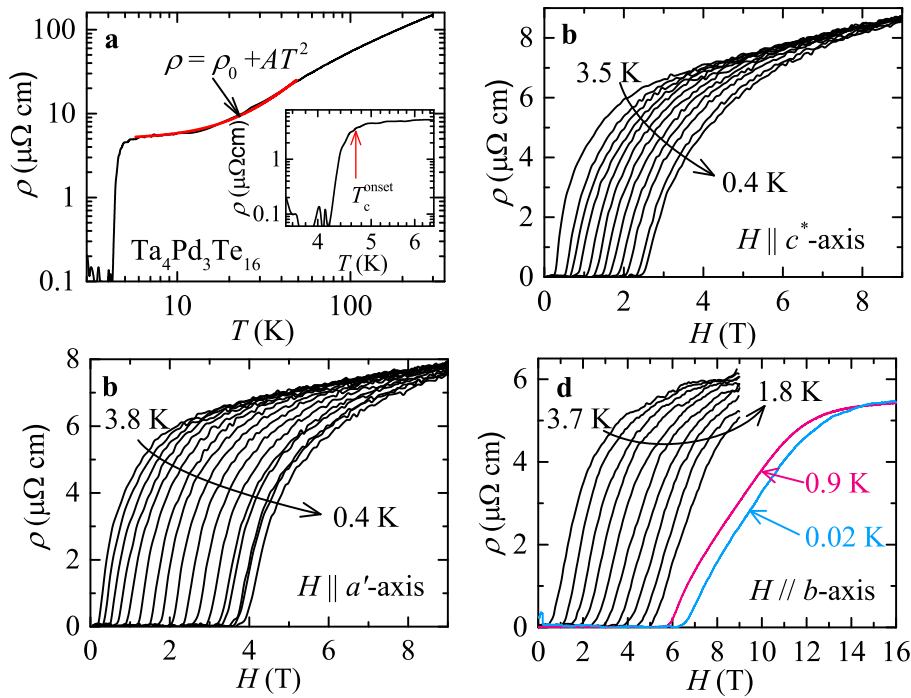}
    \caption{(a) Electrical resistivity $\rho$ of a Ta$_{4}$Pd$_{3}$Te$_{16}$ single-crystal as a function of $T$ showing an onset superconducting transition temperature $T^{\text{ons}}_{c} \sim 4.7$ K (see inset). Red line is a fit to $\rho = \rho_{0} + AT^{2}$ indicating Fermi liquid behavior. (b) $\rho$ as a function of the magnetic field $H$ applied along the $c^{\star}$-axis and for temperatures ranging from 0.4 to 3.5 K. (c) Same as in (b) but for fields along the $a^{\prime}-$axis and for $T$ ranging from 0.4 and 3.8 K. (d) Same as in (b) for fields along the $b-$axis. This panel contains data from two samples; sample \# 1 measured at temperatures ranging between 3.7 and 1.8 K (black traces) and sample \#2 measured at $T$s ranging from 0.9 to 0.02 K (colored traces).}
    \label{TPTRvH}
\end{center}
\end{figure*}
\begin{figure*}[htb]
\begin{center}
    \includegraphics[width=12 cm]{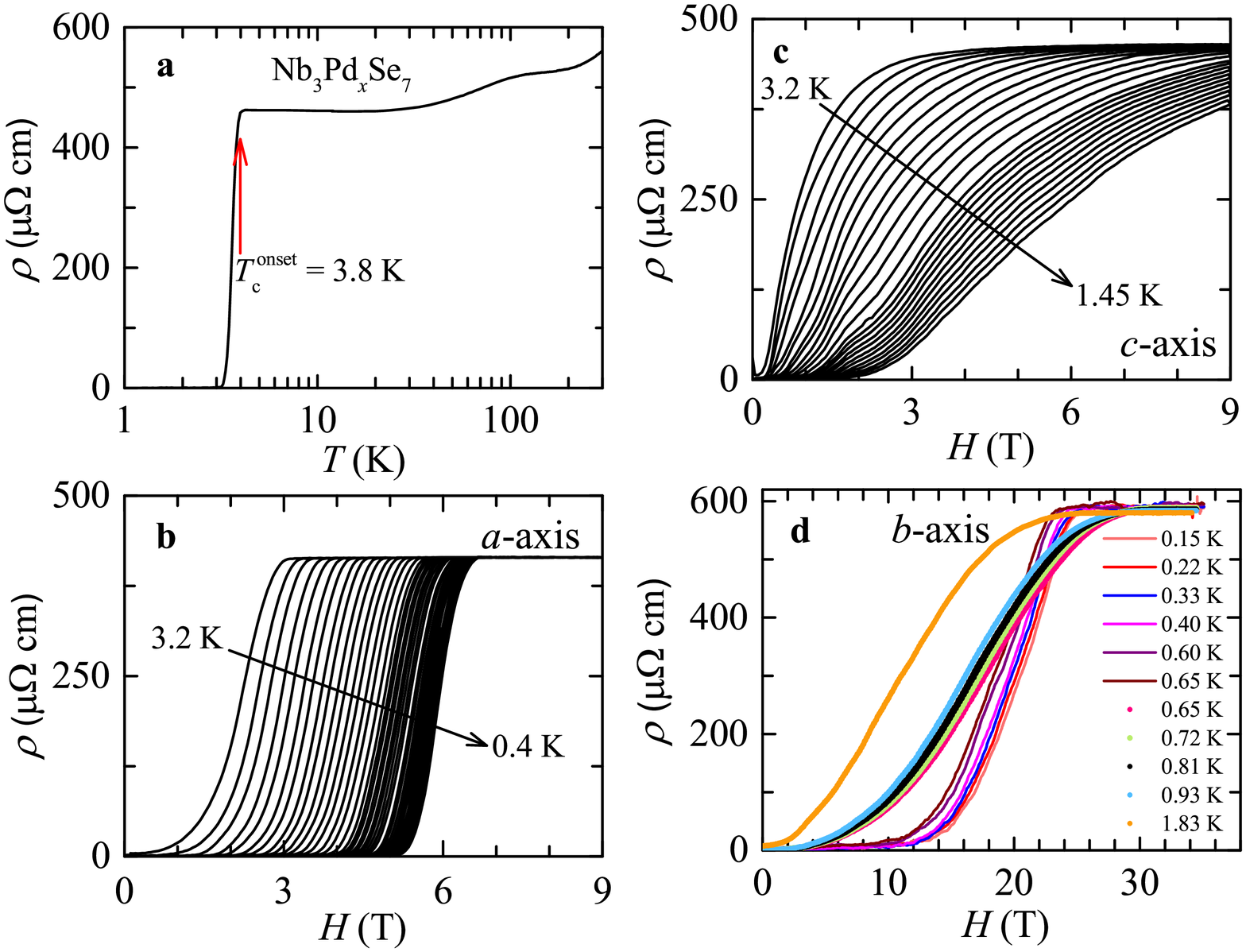}
    \caption{ (a) $\rho$ as a function of the temperature for a Nb$_3$Pd$_x$Se$_7$ single crystal displaying an onset superconducting transition temperature $T^{ons}_{c} \simeq 3.8$ K.  (b) $\rho$ as a function of the field applied along the $a-$axis and for several temperatures. (c) Same as in (b) but for fields applied along the $c-$axis. (d) Same as in (b) but for fields along the needle or the $b-$axis of the crystal. Markers and lines depict data from two different crystals having similar $T_c$s but different resistive transition widths $\Delta H = H[\rho(90 \%)]- H[\rho(10 \%)]$, where $H[\rho(10 \%)]$ corresponds to the field where $\rho$ reaches 10 \% of its value in the metallic state.}
    \label{NbPdSeResis}
\end{center}
\end{figure*}

Ta$_3$Pd$_3$Te$_{12}$ single-crystals were grown \emph{via} a self-flux method where high purity ($> 99.99$ \%) elements were mixed in the ratio Ta:Pd:Te of 2:3:15 and then sealed in an evacuated quartz ampoule. The ampoule was heated up to $950^{\circ}$ C at a rate of $15^{\circ}$ C/h and held at this temperature for 24 h. The ampoule was subsequently cooled to $650^{\circ}$ C at a rate of $5^{\circ}$ C/h from which temperature it was finally cooled down to room temperature \cite{jiao}. This procedure produces brittle single-crystals with typical dimensions of $25 \times 100 \times 1000$ $\mu$m$^{3}$. Single-crystals produced by this procedure displayed a middle-point superconducting transition $T_{c} \simeq 3.9$ K. We also followed the original synthesis method of Ref. \onlinecite{ibers} with a starting mixture of elements in the ratio Ta:Pd:Te of 2:1:11 which was heated up to $1000^\circ$ C and held at this temperature for 12 days. This procedure yielded single-crystals showing a range of $T_c$s, i.e. from 3.9 K to 4.7 K. Nb$_3$Pd$_x$Se$_7$ single-crystals were grown using the previously reported method \cite{ibers2,Alan2} showing a distribution of $T_c$s ranging from $0 - 3.5$ K which depends on the Pd content \cite{Alan2}. The stoichiometric composition was determined by energy dispersive $X$-ray spectroscopy (EDS). For the Nb$_3$Pd$_x$Se$_7$ single-crystals displaying a higher $T_c$ of  $\sim 3.5$ K, EDS indicates that the Pd content is closer to the stoichiometric value of 1, or $x \sim 0.84$, although it also reveals a considerable excess of Se, i.e. nearly 20\% in some of the crystals.
As for Ta$_4$Pd$_3$Te$_{16}$ EDS reveals samples with compositions very close to the stoichiometric values, but also samples with a Pd deficiency of $\sim 5$ \% in addition to samples apparently belonging to the Ta$_3$Pd$_3$Te$_{14}$ phase \cite{jiao2}. Electrical transport measurements were performed by using a combination of superconducting and resistive magnets, coupled to $^3$He and dilution refrigerators.

Figure 1 (a) displays the magnetic susceptibility $\chi$ as a function of the temperature $T$ for several randomly oriented Ta$_3$Pd$_3$Te$_{12}$ single-crystals. Blue markers depict $\chi(T)$ measured under zero-field cooled conditions, while red markers correspond to $\chi(T)$ measured under field cooled conditions. Both curves were acquired under a field $H= 25$ Oe. Notice the pronounced diamagnetic signal observed below $\sim 4.5$ K indicating bulk superconductivity \cite{jiao}. Bulk superconductivity is further supported by the size of the anomaly seen in the electronic contribution to the heat capacity $C_e/T$ at the superconducting transition, which is shown in Fig. 1 (b). This curve was collected from several crystals. $C_e/T$ was obtained after subtracting a $C \propto T^3$ phonon contribution term. Nevertheless, the size of the anomaly $\Delta C_e/T \simeq 22.5$ mJ/molK$^2$ relative to the electronic contribution to the heat capacity in the metallic state, i.e. $\gamma_e \simeq 25.8$ mJ/molK$^2$, is considerably smaller than the BCS value $\Delta C / \gamma_e = 1.43$. This indicates that the superconducting volume fraction is $< 1$ while also reflecting the distribution in $T_c$s among the different crystals.

Figure 2 (a) shows the resistivity $\rho$ of a Ta$_4$Pd$_3$Te$_{16}$ single-crystal for currents flowing along the $b-$axis. In contrast to both Nb$_2$Pd$_x$S$_5$ and Nb$_3$Pd$_x$Se$_7$, see Refs. \onlinecite{Alan1,Alan2}, this compound displays i) rather small residual resistivities, in the order of just a few $\mu \Omega$cm, and ii) Fermi liquid like behavior as indicated by the red line which corresponds to a fit to $\rho(T) = \rho_0 + AT^2$. Notice also that the onset of the resistive transition, i.e. the temperature where the resistivity reaches 90 \% of its value in the metallic state just above the transition, is $\sim 4.7$ K which is slightly higher than the value previously reported \cite{jiao}. Figure 2(b) displays $\rho$ as a function of the applied field $H$ applied along the $c^{\star}$-axis for several temperatures. Here, we follow the nomenclature used by Ref. \onlinecite{Pan} to identify the crystallographic axes of the single-crystal(s), with the $a^{\prime}$ and the $c^{\star}$ axes being perpendicular to the $b-$axis (and to each other) but not aligned along the crystallographic $a-$ and $c-$  axes. Figure 2(b) displays $\rho$ as a function of the applied field $H$ applied along the $a^{\prime}$-axis for several temperatures. Colored lines depict data collected from a second single-crystal having a similar $T_c$ but measured at temperatures well below 1 K.  Data from both samples collected over an extended range in temperatures are included in the phase-diagram shown below. By comparing all three panels, one can see that i) this compound is mildly anisotropic, ii) that its superconducting transition does not broaden considerably under field and iii) that the transition field does not saturate at a given value as $T$ is lowered, as one would expect for a conventional superconductor.

Subsequently, we compare the upper critical fields observed in Ta$_4$Pd$_3$Te$_{16}$ with those of Nb$_3$Pd$_x$Se$_{7}$, since both compounds belong to the same monoclinic $C2/m$ space group and display comparatively similar $T_c$s. The former compound contains heavier elements, therefore one would expect a greater influence of the spin-orbit coupling on its superconducting phase diagram. In reality, Nb$_3$Pd$_x$Se$_{7}$ single-crystals with $x \gtrsim 0.7$ and $T_c$s approaching 2 K, already display considerably higher $H_{c2}$s than those of Ta$_4$Pd$_3$Te$_{16}$ crystals with $T_c \simeq 4.7$ K.  Figure 3 (a) displays the resistivity $\rho$, for electrical currents along the \emph{b}-axis, as a function of $T$.
Similarly to our previous report this sample also displays an anomaly in the resistivity centered around 100 K of unknown origin, but the onset of its superconducting transition starts at a considerably higher temperature, i.e. $3.8$ K. Figures 3 (a) and (b) depict $\rho$ as a function of the magnetic field applied along the $a^{\prime}-$ and the $c^{\star}-$axis respectively, and for several temperatures. Notice that in contrast to Ta$_4$Pd$_3$Te$_{16}$ the values of its upper critical fields tend to saturate as $T$ is lowered, as expected for a singlet-paired superconductor. Figure 3 (c) displays $\rho$ as a function of $H$ applied along its needle or $b-$axis. In this panel we have included data from two samples, one measured under $T < 0.65$ K (solid lines) which displays a sharper transition, and a second one characterized by a broader transition under $T > 0.65$ K (solid markers).
Both samples display similar $T_c$s and hence similar values of $H_{c2}$ at intermediary temperatures, e.g. at 0.65 K. The residual resistivities were similar for both crystals hence to allow a comparison between both samples we renormalized the resistivity of one of the samples. By comparing all three panels one concludes i) that the $H_{c2}$s are anisotropic, increasing considerably for fields along the $b$-axis and ii) that the transition also broadens considerably when tilting the field from the $a-$ to the $b-$ axis suggesting a prominent role for superconducting fluctuations. Broader transitions, which are common to the two-dimensional high $T_c$ cuprates \cite{blatter}, suggest that the effective electronic dimensionality is reduced by the application of fields along the $c^{\star}-$ and more particularly along the $b-$axis as if the field renormalized the transfer integrals along crystallographic directions perpendicular to these axes.
\begin{figure*}[htb]
\begin{center}
    \includegraphics[width = 12 cm]{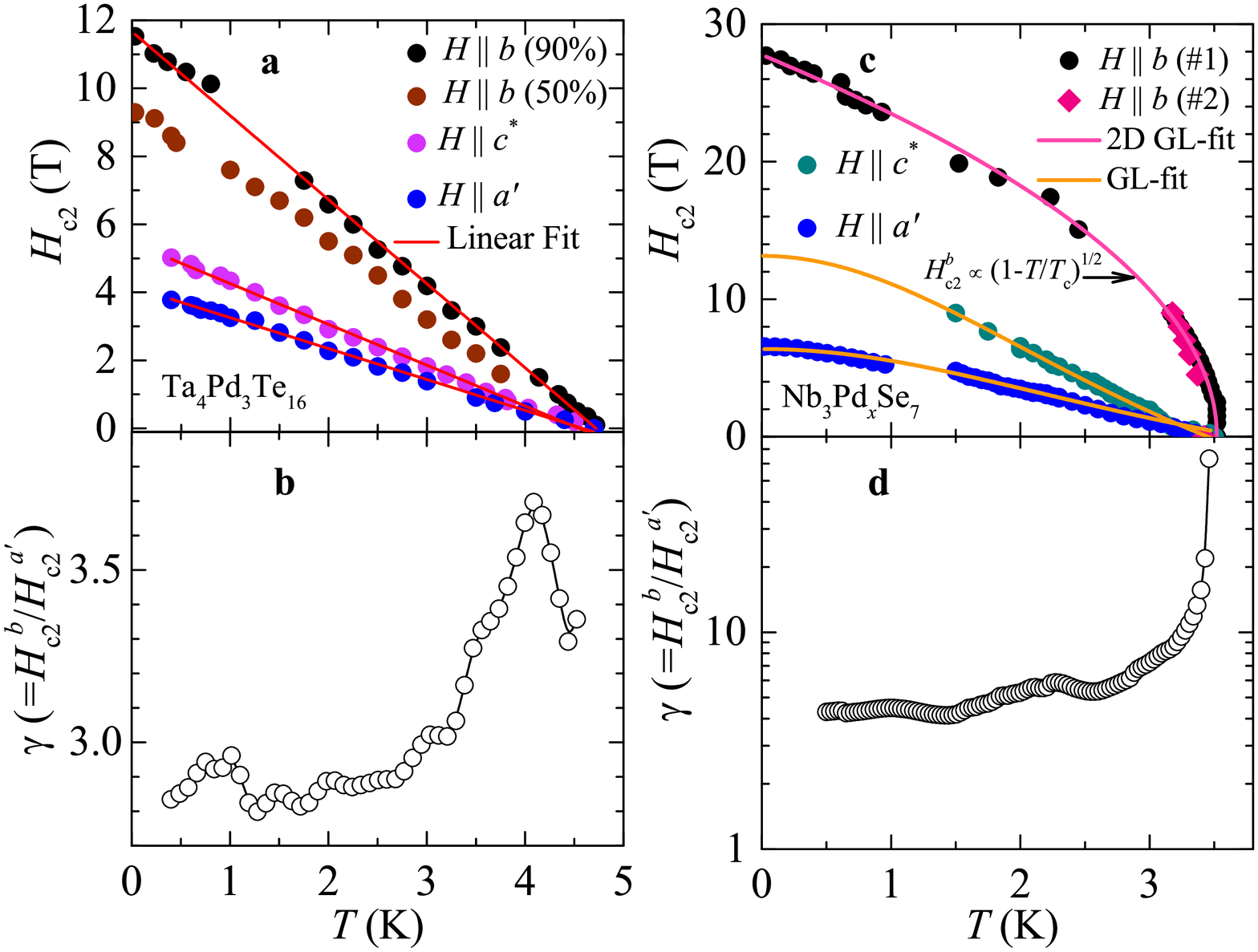}
    \caption{(a) Phase boundary between superconducting and metallic states for magnetic fields applied along all three main crystallographic axes of Ta$_{4}$Pd$_{3}$Te$_{16}$. The upper critical fields are linearly dependent in $T$ for all three crystallographic orientations showing no sign of saturation. For fields
    applied along the $b-$axis, $H_{c2}^b (T \rightarrow 0 \text{ K})$ surpasses the weak coupling Pauli limiting value $H_p = 1.84 \times T_c \simeq 8.5 $ T. (b) Superconducting anisotropy $\gamma = H_{c2}^b/ H_{c2}^{a^{\prime}}$, where both variables correspond to fields where the resistivity reaches 90 \% of its value in the metallic state. (c) Superconducting phase diagram for Nb$_{3}$Pd$_{x}$Se$_{7}$ single crystals having a middle point resistive transition at $T_c \simeq 3.5$ K. Notice, the much larger values of $H_{c2}$ relative to the Ta compound, as well as its anomalous phase-boundary for fields along the \emph{b}-axis. Magenta line corresponds to a fit to the expression $H_{c2}^b \propto (1-t)^{1/2}$ where $t = T/T_c$, which describes a nearly two-dimensional superconductor in the vicinity of $T_c$.
    (c) Superconductivity anisotropy $\gamma = H_{c2}^b/ H_{c2}^{a^{\prime}}$ for Nb$_{3}$Pd$_{x}$Se$_{7}$. (\textbf{d}) The superconductivity anisotropy for Nb$_{3}$Pd$_{x}$Se$_{7}$ between the $b$-axis and the $a^{\prime}$-axis.}
    \label{TPTPD}
\end{center}
\end{figure*}
\begin{figure*}[ht]
\begin{center}
    \includegraphics[width=12cm]{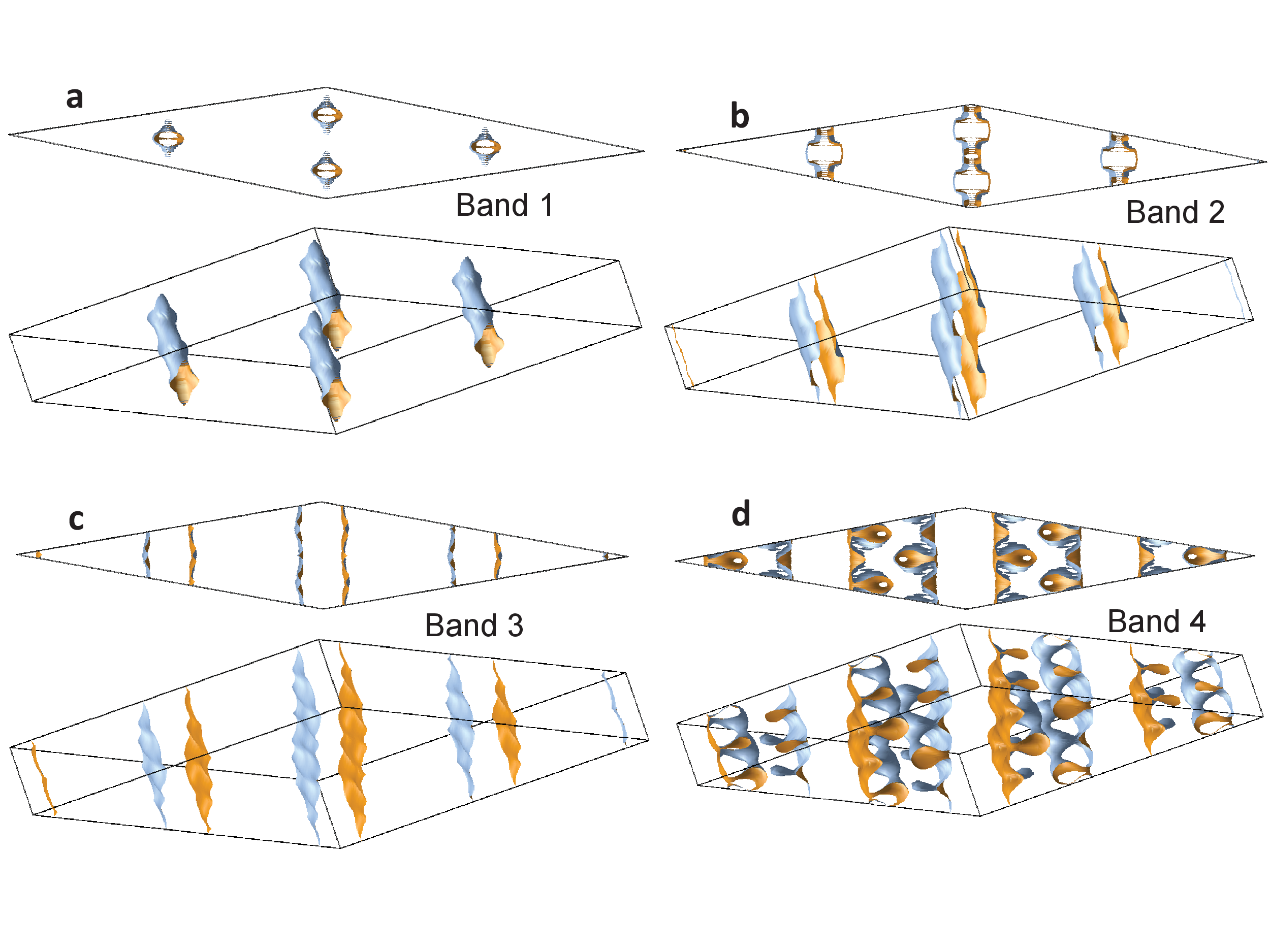}
    \caption{Fermi surfaces of Ta$_{4}$Pd$_{3}$Te$_{16}$ according to density functional theory calculations. Four bands intersect the Fermi level leading to (a) two-dimensional cylinders (band 1), in (b) and (c) quasi-one-dimensional Fermi surface sheets (resulting from bands 2 and 3), and (c) three dimensional anisotropic networks resulting from band 4. Each set of Fermi surfaces is depicted through two different perspectives, perpendicularly to the $c^{\star}-$axis and through a side view of the monoclinic Brillouin-zone.}
    \label{FSs}
\end{center}
\end{figure*}

In Figure 4 we plot the superconducting phase diagrams and concomitant superconducting anisotropies for both compounds. As seen in Fig. 4 (a), the upper critical fields for Ta$_4$Pd$_3$Te$_{16}$ follow a simple linear dependence on temperature over the \emph{entire} $T$ range. Here, for all three field orientations we used the 90 \% criteria, or the onset of the resistive transition in order to avoid any possible influence from phases related to vortex matter. Nevertheless, for fields along the needle-axis we also included points corresponding to the middle point of the resistive transition (brown markers), or the 50 \% criterium, to indicate that such anomalous linear dependence cannot be attributed to superconducting fluctuations. For $H \parallel b-$axis, one observes non-saturation of $H_{c2}^b$ when $H$ approaches or surpasses the value of the Pauli limiting field
$H_p \simeq 1.84 \times \sqrt{1+ \lambda} \times T_c$ which in the weak coupling regime, or for $\lambda \sim 0$, leads to $1.84 \times 4.7 \sim 8.7$ T which is smaller than the measured value. Here, $\lambda$ reflects the strength of the electron-phonon or of the electron-boson coupling. This observation contrasts with what is expected for singlet pairing;
for dirty and nearly isotropic superconductors (single crystal \emph{x}-ray diffraction indeed indicates a sizeable amount of site disorder) the influence of a reduced mean free path is usually described through the Maki-de Gennes relation \cite{maki}:
\begin{equation}
 \ln(T/T_c) = \psi (1/2 + eDH_{c2}/ 2\pi k_B T) - \psi (1/2)
\end{equation}
where $\psi$ is the digamma function and $D = \pi^2 k^2_B/ (3e^2 \gamma \rho_0)$ the diffusion constant,
with $\gamma$ being the electronic coefficient in the heat-capacity and $\rho_0$ the residual resistivity.
In contrast to what is seen here, such expression leads to a saturation of the upper-critical fields at temperatures
well below $T_c$ see, for example, Ref. \onlinecite{schuller}. In Fe pnictide superconductors one observes a similar linear
dependence of $H_{c2}$ on $T$ for fields along certain orientations, but not for \emph{all} orientations,
which was claimed to result from the orbital limiting effect \cite{singleton}. A linear dependence
of $H_{c2}$ on $T$ was also observed in few layered transition metal dichalcogenides for fields applied
perpendicularly to the conducting planes \cite{Mak,Iwasa} and described in terms of a phenomenological two-dimensional Ginzburg-Landau model:
\begin{equation}
H_{c2}(T) = \frac{\phi_0}{2 \pi \overline{\xi}(0)^2} \left(1- \frac{T}{T_c} \right)
\end{equation}
where $\phi_0$ correspond to the quantum of flux and $\overline{\xi}(0)$ to an average in-plane
Ginzburg-Landau coherence length at $T = 0$ K. Nevertheless, it is seemingly unphysical for a bulk single-crystal
to display two-dimensional superconductivity for every orientation of an external magnetic field.
The contrast in phase-diagrams between Ta$_4$Pd$_3$Te$_{16}$ and Nb$_{3}$Pd$_{x}$S$e_{7}$, with both compounds being characterized by disorder
and as seen below multi-band superconductivity, is at odds with the scenario proposed by Ref. \onlinecite{Singh1} to explain this linear dependence on $T$.
Instead, it might indicate that Ta$_4$Pd$_3$Te$_{16}$ is orbital-limited over the entire temperature range,
or equivalently that the Pauli limiting field has been renormalized to higher values due to correlations, i.e. $\lambda > 1$
as seen in strongly coupled superconductors. This is supported by a Ginzburg-Landau analysis of $H_{c2} (T)$ as a function of $(1-T/T_c)$ in
the vicinity of $T_c$ (similar plots and related discussion can be found in Refs. \onlinecite{Alan1,Alan2})
which yield unreasonably high values for the Pauli limiting field. Finally, the mild dependence on temperature of the superconducting anisotropy
$\gamma = H_{c2}^b/ H_{c2}^{a^{\prime}}$ as seen in Fig. 4 (b) suggests multi-band superconductivity as previously argued \cite{Alan1,Alan2} for
Nb$_{2}$Pd$_{x}$S$_{5}$ and Nb$_{3}$Pd$_{x}$S$e_{7}$. At the lowest temperatures a $\gamma \simeq 3$  is comparable to
$\gamma \simeq 4$ previously obtained for Nb$_{3}$Pd$_{x}$S$e_{7}$.

For Nb$_{3}$Pd$_{x}$S$e_{7}$ single-crystals, with a middle point $T_c \simeq 3.5$ K, the phase-boundary between the superconducting and the metallic states for fields applied perpendicularly to the needle-axis behave quite similarly to those displaying a $T_c \simeq 1.8$ K, namely a linear dependence on $T$ followed by saturation of the $H_{c2}$s at the lowest temperatures. Roughly, the $H_{c2}$s between both sets of samples scale linearly with $T_c$. Notice how for both sets of samples, and for fields nearly along the $c^{\star}$-axis, $H_{c2}^{c^{\star}}$ for $T \sim 0.5 \times T_c$ already surpasses the Pauli limiting value. Orange lines are fits to the Ginzburg-Landau expression:
\begin{equation}
 H_{c2}^{a^{\prime}, c^{\star}}(T) = \frac{\phi_0}{2 \pi \xi_b(0) \xi(0)_{c^{\star}, a^{\prime}}} \left(\frac{1-t^2}{1+t^2} \right)
\end{equation}

where $t=T/T_c$ is the reduced temperature and $\xi(0)_{a^{\prime}, b, c^{\star}}$ is the coherence length along the $a^{\prime}$, or $b$, or $c^{\star}$ axis. As seen, $H_{c2}^{a^{\prime},c^{\star}}$s for fields applied perpendicularly to the needle-axis, can be well-described by conventional behavior, albeit leading to $H_{c2}(T\rightarrow 0 \text{ K}) \sim 2 \times H_p$ for fields applied along the $c-$axis. The observed saturation in $H_{c2}$s seen at the lowest $T$s clearly indicates that Nb$_{3}$Pd$_{x}$S$e_{7}$ is a Pauli limited superconductor. In contrast, for fields along the $b-$axis we observe a fast increase of $H_{c2}$ as $T$ is lowered with respect to $T_c$. We observed this behavior in 3 crystals with $T_c$s approaching 3.5 K; for both samples whose raw data are displayed in Fig. 3 (d) and which were used to build the extended phase-boundary (black markers) shown in Fig. 4(c), and a third sample (diamonds) measured in the vicinity of $T_c$. Notice that for this orientation $H_{c2}^b$ does \emph{not} saturate but increases as the $T$ is lowered, contrasting markedly with what is observed for fields along the $a^{\prime}$-axis. Magenta line is a fit to the phenomenological two-dimensional Ginzburg-Landau expression used to describe $H_{c2}^{\parallel}$ for fields along the planes of ultra-thin crystals of transition metal dichalcogenides \cite{Iwasa}:
\begin{equation}
 H_{c2}^{b}(T) = \frac{\phi_0\sqrt{12}}{2 \pi \overline{\xi}(0)d_{sc}} \sqrt{1-T/T_c}
\end{equation}

where $d_{sc}$ corresponds to the effective thickness of the superconducting layers.
As discussed in Ref. \onlinecite{Mak} this expression can be derived from Eq. (1) in the vicinity of $T_c$ by modifying it to describe not the pair-breaking effect produced by impurities but the Pauli pair breaking effect. Remarkably, this expression fits the data over the entire $T$ range, indicating that for fields along the $b-$axis Nb$_{3}$Pd$_{x}$S$e_{7}$ behaves is if it was a two-dimensional superconductor, or as if the spins were locked at a direction perpendicular to the $b$-axis through a strong spin-orbit coupling \cite{Iwasa,Mak,fedor}. Further evidence for two-dimensional superconductivity is provided by the superconducting anisotropy $\gamma$ which seemingly ``diverges" as one approaches $T_c$, while displaying at the lowest $T$s the same value $\gamma \simeq 4$ previously observed in the lower $T_c$ samples. For $H^b_{c2}(0 \text{ K}) \sim 28$ T, one obtains $\overline{\xi}(0)d_{sc} = 4071$ \AA$^2$. If one assumed that the effective thickness of the superconducting slabs was comparable to the lattice constant along the $c-$axis \cite{Alan2}  or 21.0370 {\AA}, one would obtain that $\overline{\xi}(0)$ would extend over $\sim 15$ lattice constants along the $a-$axis.

\section{discussion}
\begin{figure*}[htb]
\begin{center}
    \includegraphics[width=13cm]{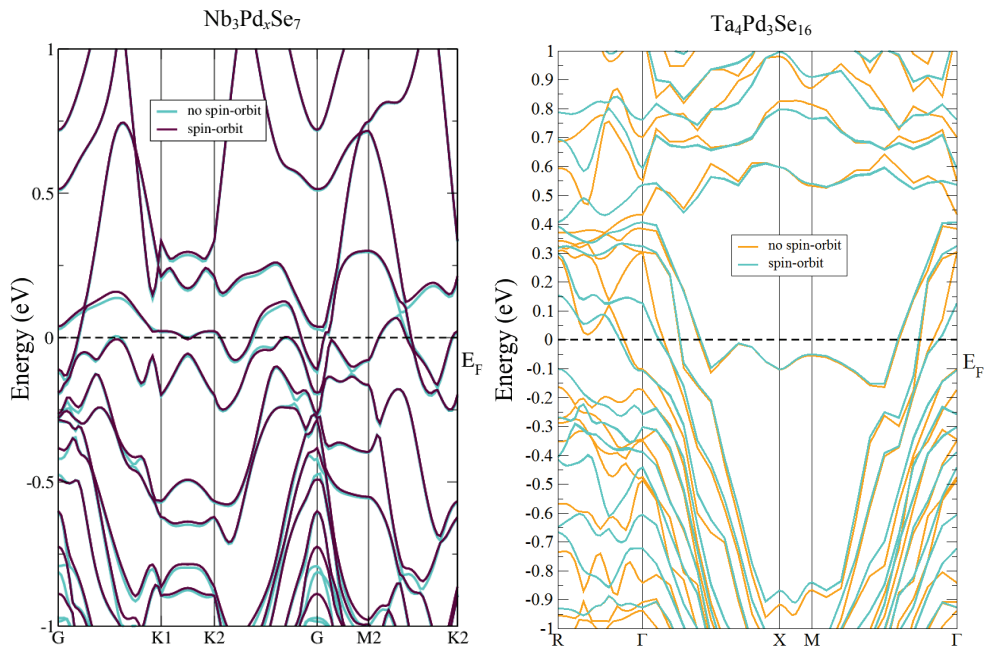}
    \caption{Left panel: Band structure of Nb$_3$Pd$_x$Se$_7$ in absence (black lines) and with the inclusion (cyan lines)of the spin-orbit coupling. Right panel: Band structure of Ta$_4$Pd$_3$Se$_{16}$ in absence (orange lines) and with the inclusion (cyan lines) of the spin-orbit coupling.}
    \label{FSs}
\end{center}
\end{figure*}

The Fermi surface of all of the Pd-chalcogenide based superconductors is quite similar, that is composed of quasi-one-dimensional and two-dimensional sheets along with three-dimensional Fermi surface networks, see Fig. 5 for the Fermi surface of Ta$_4$Pd$_3$Te$_{16}$ resulting from the density functional theory (DFT) calculations discussed below, and also Refs. \onlinecite{Alan1,Alan2,Singh1,Singh2,Singh3}. Therefore, it would be quite difficult to explain the marked differences between the phase-diagrams of Nb$_3$Pd$_x$Se$_7$ and Ta$_4$Pd$_3$Te$_{16}$ based solely on their similar crystallography and electronic structures at the Fermi level. However, and as previously mentioned, in transition metal dichalcogenides the strong spin-orbit interaction is known to split the valence and the conduction bands aligning or ``locking" the spins of the charge carriers along a direction perpendicular to the conducting planes with each spin-orbit split sub-band developing a spin texture, namely spins pointing in opposite directions at the $K$ and $-K$ valleys. This moderately large, and valley-dependent, Zeeman-like spin splitting in the vicinity of the $K$ points, would protect singlet Cooper pairing among carriers originally located at the $K$ and $-K$ valleys: each carrier would have a locked, out of the plane spin polarization of opposite polarity. This inter-valley Ising like pairing is claimed, by the authors of Refs. \onlinecite{Mak,Iwasa}, to enhance their upper critical fields well beyond the BCS Pauli limiting field.

Perhaps a similar scenario might be applicable to Nb$_3$Pd$_x$Se$_7$ and also to the Nb$_2$Pd$_x$(Se,S)$_5$ series, namely Ising pairing among carriers that have their spins locked along a direction perpendicular to the $b-$axis, probably along the $a$ or $a^{\prime}$-axis (since both systems display the lowest $H_{c2}$s along this orientation), due to strong spin-orbit coupling. Similarly to the transition metal dichalcogenide single-layers, this would explain the extremely large values of $H_{c2}^b$ observed in the Pd based superconductors, which far exceed their Pauli limiting field. Such scenario might explain why Nb$_3$Pd$_x$Se$_7$ would have a relatively higher value of $H_{c2}^b$ in the limit of zero temperature, i.e. $\sim 28$ T versus $\sim 37.5$ T for Nb$_2$Pd$_x$S$_5$, given their relative $T_c$s of $\sim 3.5$ K and $\sim 6.5$ K, respectively. Nb$_3$Pd$_x$Se$_7$ has a slightly higher content of Nb relative to the chalcogen element with Se being heavier than S thus leading to a potentially stronger spin-orbit coupling. In this scenario, if Ta$_4$Pd$_3$Te$_{16}$ was not an orbital limited superconductor characterized by higher values of $\xi(0)_{a^{\prime}, b, c^{\star}}$ relative to Nb$_3$Pd$_x$Se$_7$, it should have displayed even higher values of $H_{c2}^b$. In this respect, and in order to understand the role of spin-orbit coupling, we evaluated the effect of SO-coupling on the band structure of both compounds through DFT calculations using the Wien2K implementation. We find that the bands of Ta$_4$Pd$_3$Te$_{16}$ are in general more dispersive and, not surprisingly, exhibit stronger SO-splitting than those of Nb$_3$Pd$_x$Se$_7$, as can be seen in Fig. 6.  There is a relatively flat band in Ta$_4$Pd$_3$Te$_{16}$ that crosses the Fermi level; it has primarily Pd-character hybridized with Te and might therefore be expected to exhibit reasonably strong spin-orbit effects due to the two heavy elements, see Fig. 6. However, our calculations with and without the SO term show that there is nearly no effect at all on this band, though SO-splitting can be seen in other (Ta/Te-derived) bands and cause small changes in the Fermi surfaces.  In Nb$_3$Pd$_x$Se$_7$, on the other hand, the effects of the spin-orbit are much smaller and are not even visible to the eye at the Fermi energy, despite the heavy Pd-character of most bands crossing $E_F$. Therefore, the Ising-like pairing scenario driven by SO-coupling does not seem to be appropriate for the superconducting state of the Pd chalcogenides. In addition, by analyzing Fig. 6 one realizes that one cannot find, for either compound, two nearly degenerate minima (or ``valleys") in the conduction band(s) (or two nearly degenerate maxima in the valence band(s)) which should, in the Ising pairing scenario, be located at high symmetry points and provide the charge carriers for pairing.

From the perspective of the proximity to magnetism both compounds behave differently. First, the Nb$_3$Pd$_x$Se$_7$
is already near a magnetic instability.  In our calculations, a spin-polarized state with small moments on
some of the Nb ions is lower in energy than a spin-unpolarized state, though an ordered state is not observed in
experiment.  This kind of discrepancy is often a sign of disordered local moments (paramagnetism).  In
Ta$_4$Pd$_3$Te$_{16}$, the energetic ground state is fully non-magnetic.

In addition to the standard spin-orbit coupling term, we can also gauge the response to an external magnetic field by adding
a term of the form $(V_{B_{\text{ext}}} = \mu_B \vec{B_{\text{ext}}}\cdot (\vec{l}+\vec{s}))$ to the potential,
where $\vec{l}$ is the angular momentum and $\vec{s}$ the spin.  This field is applied only within spheres drawn
around each constituent ion (consistent with our muffin-tin methodology) and the resulting orbital moments are
calculated within the same sphere.  Similarly, a spin moment is calculated within the sphere, but because
polarization of the charge does not require an evaluation of orbital occupancy, we can also gauge the spin
response of the interstitial charge, \emph{i.e.} that charge which cannot be assigned to any given site.  We apply
this field along three directions: along the \emph{b}-axis (``\emph{b}-axis"), perpendicular to the \emph{b}-axis along the Nb-Se chain direction
(``chain") and perpendicular to the \emph{b}-axis along the interchain direction (``interchain").

In Nb$_3$Pd$_x$Se$_7$, there is a clear spin response and an anisotropic orbital response to the field.
Regardless of field direction, the Nb(1) and Nb(2) atoms carry a local spin moment of $\sim 0.4$ $\mu_B$
and the Nb(3) atom that sits nearest the Pd(2) sites carries almost no moment at all.  The Se ions are
spin- and orbital-unpolarized.  The Pd sites show an intriguing anisotropy in orbital moment polarization.
Both Pd sites have a \emph{C2/m} square planar symmetry, and each site shows a much stronger tendency toward
orbital polarization when the field points perpendicular the plane of the square.  For Pd(1) the easy field
direction is the ``\emph{b}-axis" direction where an orbital moment of $\sim 0.2$ $\mu_B$ forms in response to the
applied field, compared to $\sim 0.1$ $\mu_B$ in the other two directions.  The Pd(2) sites are oriented
perpendicularly such that the easy field direction is approximately along the ``chain" direction.  Both Pd
sites carry a spin moment of less than 0.1 $\mu_B$ with no noticeable anisotropy as a function of field
direction. The increase in orbital polarization as a function of the field is likely to pin the spin moment 
to the lattice and hence contribute to the high upper critical fields.

In contrast, the overall response of Ta$_4$Pd$_3$Te$_{16}$ to the applied field is very different from that of
Nb$_3$Pd$_x$Se$_7$: none of ions carry a significant spin moment and the orbital moment on Pd is almost
entirely anisotropic.  The interstitial spin moment is very close to that of Nb$_3$Pd$_x$Se$_7$ at $\sim
0.41$ $\mu_B$/f.u. and the orbital moments on the Ta ions are approximately $\sim 0.2$ $\mu_B$ (identical
to the Nb).  The Pd ions have a spin moment of $\sim 0.06$ $\mu_B$ and an orbital moment of $\sim 0.11$
$\mu_B$, almost identical to the values found for Pd in Nb$_3$Pd$_x$Se$_7$.  However, the orbital response
of the Pd ions in Ta$_4$Pd$_3$Te$_{16}$ is insensitive to the field direction.  It should be noted that
there is a slight tendency for the interstitial charge to spin-polarize more strongly when the field is
along the $c$-axis, with the moment growing to $0.52$ $\mu_B$/f.u.

The smaller spin-orbit coupling, particularly near the Fermi energy, combined with the larger spin response to the application of a magnetic field,
suggests that Nb$_3$Pd$_x$Se$_7$ might have a smaller critical field ratio than Ta$_4$Pd$_3$Te$_{16}$, and yet we see precisely
the opposite in our experiments. Hence, we are led to conclude that the strength of the spin-orbit coupling and proximity to
magnetism is of little relevance for the superconducting phase-diagram of Ta$_4$Pd$_3$Te$_{16}$ further supporting the
orbital-limiting scenario over the entire temperature range.

It is tempting to tie both the variation in the superconducting temperature as a function of Pd content and the anisotropy and
strength of $H_{c2}$ to the strongly anisotropic (2D) orbital response of the Pd ions. To gauge the effect of changing Pd content
on the system, we doubled the unit cell and removed one of the Pd(2) atoms from the compound.  This results in
Nb$_3$Pd$_{0.5}$Se$_7$, a lower Pd content than our model compound Nb$_3$PdSe$_7$ and also lower than the observed stoichiometry
of Nb$_3$Pd$_{0.84}$Se$_{7+\delta}$.  Nonetheless, we believe that the trends seen in this calculation should mirror the effects
of variation of the Pd(2) content in real materials.  Interestingly, very little changes overall with the decreased Pd content.
The previously unpolarized Nb(3) ion gains a small amount of spin polarization, while all other Nb, Pd(1), and Se spin and
orbital moments are indistinguishable from the original (high Pd content) calculation.  The remaining Pd(2) ion, however, loses
nearly all of its spin polarization, but retains most of the orbital polarization.

Two-dimensional superconductivity in Nb$_3$Pd$_{0.84}$Se$_{7+\delta}$ could result from weakly-coupled superconducting planes
which are themselves formed by chains composed of square-planar and trigonal-prismatic Se polyhedra approximately centered around
the Pd and the Nb atoms, see Fig. 3 in Ref. \onlinecite{Alan2}. These planes are connected just by square coordinated Pd(2)
atoms, whose molar fraction is non-stoichiometric. In addition, one would expect that the excess Se $\delta$ became interstitial
Se randomly placed in between adjacent planes acting perhaps as pair-breaking impurities. Unconventional superconductors are
particularly susceptible to non-magnetic impurities \cite{andy}.  Hence, given the experimental evidence, we speculate that the
Cooper pair coherence length is rather anisotropic but becomes comparable to the inter-planar distance at low $T$s. Nevertheless,
we do not have an explanation for the lack of saturation in the $H_{c2}^b$s of Nb$_3$Pd$_x$Se$_7$ which is observed in the
\emph{same} compound \cite{Alan2} when it displays lower $T_c$s.  It is tempting to attribute this lack of saturation to a possible lack of local inversion symmetry, which is indeed observed on the Pd sites of Ta$_4$Pd$_3$Te$_{16}$ which displays a $C1$ instead of the $C2/m$ symmetry. Lack of inversion symmetry, which leads to an asymmetric spin-orbit coupling, was claimed to minimize the role of the Pauli limiting effect in spin-singlet superconductors \cite{frigeri} and even to an admix of spin-singlet and spin-triplet superconducting pairing channels \cite{yuan}. However, the Nb compound preserves inversion symmetry while the contrasting behavior as a function of the Pd stoichiometry suggests instead a possible change in the superconducting pairing symmetry probably triggered by the rapid increase in superconducting anisotropy. The correlation of
$T_c$ with the decrease in superconducting and/or electronic anisotropy is difficult to reconcile with conventional $s-$wave
pairing. Insofar, contradictory evidence for unconventional pairing has only been reported for the Ta$_4$Pd$_3$Te$_{16}$ compound
\cite{HHWen,Pan,fan}, but a similar crystalline space group and an akin electronic structure suggests that unconventional
superconductivity is a possibility for all of the Pd and chalcogenide based superconductors. Unconventional superconductivity is
further supported by the anomalous temperature dependence of the upper critical fields reported in this manuscript.


\section{Acknowledgements}
We acknowledge T. Besara and T. Siegrist for the \emph{X}-ray diffraction
measurements on the Ta$_4$Pd$_3$Te$_{16}$ single-crystals.
The NHMFL is supported by NSF through NSF-DMR-1157490 and the
State of Florida.  L.~B. is supported by DOE-BES through award DE-SC0002613.


\end{document}